# ADMM Penalty Parameter Evaluation for Networked Microgrid Energy Management

Jesus Silva-Rodriguez and Xingpeng Li
Department of Electrical and Computer Engineering
University of Houston
Houston, TX
jasilvarodriguez@uh.edu

*Abstract*—The alternating direction method of multipliers (ADMM) is a powerful algorithm for solving decentralized optimization problems including networked microgrid energy management (NetMEM). However, its performance is highly sensitive to the selection of its penalty parameter $\rho$, which can lead to slow convergence, suboptimal solutions, or even algorithm divergence. This paper evaluates and compares three district ADMM formulations to solve the NetMEM problem, which explore different methods to determine appropriate stopping points, aiming to yield high-quality solutions. Furthermore, an adaptive penalty heuristic is also incorporated into each method to analyze its potential impact on ADMM performance. Different case studies on networks of varying sizes demonstrate that an objective-based ADMM approach, denominated as OB-ADMM, is significantly more robust to the choice of $\rho$, consistently yielding solutions closer to the centralized optimal benchmark by preventing premature algorithm stopping.

## Nomenclature

*Sets*
- $M$: Local microgrids in the network.
- $G$: Generator units.
- $E$: Energy storage units.
- $T$: Time intervals.

*Parameters*
- $C^{SU}$: Start-up cost of a generator.
- $C^{NL}$: No-load cost of a generator.
- $C^G$: Operating cost of a generator.
- $P^{net}$: Net load demand supplied by each microgrid.
- $C^{grid+}$: Grid power import price.
- $C^{grid-}$: Grid power export price.
- $C^{Np}$: Price of power exchanged among microgrids.
- $P^E_{lim}$: Microgrid tie-line limit to central node.
- $\rho$: ADMM penalty parameter.
- $\beta$: Average objective value rate of change threshold.
- $k_s$: Algorithm iterations offset.

*Variables*
- $P^G$: Generator power output.
- $u^G$: Generator binary status indicator.
- $v^G$: Generator binary start-up indicator.
- $P^{ESc}$: Energy storage unit power input.
- $P^{ESd}$: Energy storage unit power output.
- $P^{grid+}$: Power import from main grid.
- $P^{grid-}$: Power export from main grid.
- $P^{N+}$: Power import into one microgrid from another.
- $P^{N-}$: Power export out of one microgrid to another.
- $P^E$: Total power exchange of a single microgrid
- $u^{N+}$: Microgrid binary power import indicator.
- $u^{N-}$: Microgrid binary power export indicator.
- $y^L$: ADMM Lagrange multiplier
- $r^L$: Primal residual.
- $s^L$: Dual residual.
- $\varepsilon$: Solution feasibility.

## I. Introduction

A microgrid (MG) is a viable solution for the integration of distributed energy resources (DER) as well as for resilience enhancement of the bulk power system thanks to their ability to operate islanded from the main grid in case of emergencies [1]-[2]. Moreover, a network of such MG systems has the ability to further enhance such benefits, particularly pertaining to the integration of renewable energy resources (RES) by allowing multiple MGs to have access to a wider variety of resources from different locations, improving energy flexibility [2].

However, when interconnecting multiple MG agents into a network and performing a consolidated energy management strategy for all network participants, a decentralized optimization approach is needed to facilitate modern distributed control of a system, eliminate the role of a central control coordination, increase reliability and robustness, and preserve the privacy and autonomy of each MG network participant [3].

Since decentralized optimization is preferred to achieve the previously mentioned benefits, a distributed optimization method can be implemented such as the alternating direction method of multipliers (ADMM), which is a simple but powerful algorithm for convex optimization problems that can be decomposed into several local subproblems and coordinate their optimization globally via specific constraints that interrelate the different subproblems [4]. ADMM has already been implemented in different ways to solve the networked microgrid energy management (NetMEM) problem where multi-MG network operations have been decentralized by implementing different formulations spanning from problem convexification to multi-objective approaches to ensure ADMM can be used effectively, as well as different methods to study and ensure appropriate algorithm convergence to a global optimal solution [5]-[10].

One of the drawbacks and a recurring challenge found in different implementations of ADMM is its sensitivity to the choice of a single hyperparameter known as the penalty on which rate of convergence is strongly dependent. Careful selection of this parameter must be made to obtain adequate convergence rate, otherwise a suboptimal solution may be obtained or even algorithm divergence may occur [11]-[13]. There are different heuristics that have been implemented to attempt to improve ADMM convergence despite penalty parameter selection, or adjust the penalty value dynamically to ensure adequate solutions near the optimum for different scenarios and applications.

This paper will focus on three different ADMM formulations tailored to the NetMEM problem. These formulations are the standard ADMM formulation [4]-[5], [11], a relaxed ADMM which solves a version of the NetMEM with relaxed binary variables first, followed by a refinement step solving the original problem [14], and an objective-based approach that incorporates additional criteria to check for convergence based on optimality performance [15]-[16]. In addition, for each of these three methods, an adaptive penalty step [6], [12] will be added as well to analyze the potential benefit of such modification to the ADMM algorithm for the NetMEM problem. These methods have been implemented on different problem types. However, the optimization concepts are very similar to those of the NetMEM problem, which will be proven to be effective for this case. Moreover, a comparison across the different methods reveals which ADMM formulation is more suitable for networks of MGs.

## II. NETMG PROBLEM FORMULATION

A simplified formulation for a small-scale interconnected network of MGs is used in this paper to implement and test different distributed optimization methods via ADMM. This formulation minimizes the combined operation cost of all MGs participating in the network, defined by (1).

$$minimize \ \sum_{m \in M} f_{cost,m} \quad (1a)$$

$$f_{cost,m} = \sum_{t \in T} \{ \sum_{g \in G} [SU_{m,g}^G v_{m,g,t}^G + \Delta t \cdot (NL_{m,g}^G u_{m,g,t}^G + C_{m,g}^G P_{m,g,t}^G)] + \Delta t \cdot [C_t^{grid+} P_{m,t}^{grid+} - C_t^{grid-} P_{m,t}^{grid-} + C_t^{Np} \sum_{n \in M, n \neq m} (P_{m,n,t}^{N+} - P_{m,n,t}^{N-})] \} \quad (1b)$$

The formulation is subject to several local and global constraints. Each MG's generators and energy storage units adhere to standard operational constraints that regulate power outputs, charging and discharging limits, start-up and shutdown logic, etcetera [17]. Moreover, unique constraints to regulate the power exchange of every microgrid with the rest of the network tailored for a central node topology are defined in (2), including power exchanges with the main grid, limited by the same tie-line. Additionally, constraint (3) represents the power balance of each MG, and constraint (4) serves as the main global constraint that interconnects the operations of all MGs in the network. This global constraint practically ensures feasibility of the global solution.

$$P_{m,t}^E = P_{m,t}^{grid+} - P_{m,t}^{grid-} + \sum_{n \in M, n \neq m} (P_{m,n,t}^{N+} - P_{m,n,t}^{N-}), \forall m \in M, t \in T \quad (2a)$$

$$-P_{lim,m}^E \leq P_{m,t}^E \leq P_{lim,m}^E, \forall m \in M, t \in T \quad (2b)$$

$$0 \leq P_{m,n,t}^{N+} \leq P_{lim,m}^E u_{m,t}^{N+}, \forall m, n \in M, n \neq m, t \in T \quad (2c)$$

$$0 \leq P_{m,n,t}^{N-} \leq P_{lim,m}^E u_{m,t}^{N-}, \forall m, n \in M, n \neq m, t \in T \quad (2d)$$

$$0 \leq P_{m,t}^{grid+} \leq P_{lim,m}^E u_{m,t}^{N+}, \forall m \in M, t \in T \quad (2e)$$

$$0 \leq P_{m,t}^{grid-} \leq P_{lim,m}^E u_{m,t}^{N-}, \forall m \in M, t \in T \quad (2f)$$

$$u_{m,t}^{N+} + u_{m,t}^{N-} \leq 1, \forall m \in M, t \in T \quad (2g)$$

$$\sum_{g \in G} P_{m,g,t}^G + \sum_{b \in ES} (P_{m,e,t}^{ESd} - P_{m,e,t}^{ESc}) + P_{m,t}^{grid+} - P_{m,t}^{grid-} + \sum_{n \in M, n \neq m} (P_{m,n,t}^{N+} - P_{m,n,t}^{N-}) = P_{m,t}^{net}, \forall m \in M, t \in T \quad (3)$$

$$P_{m,n,t}^{N+} = P_{n,m,t}^{N-}, \forall m, n \in M, n \neq m, t \in T \quad (4)$$

This problem formulation for the NetMEM is in a centralized form which collectively optimizes the network looking for a global minimum objective operation cost. The next section shows how this problem is decentralized to preserve each MG's autonomy via ADMM.

## III. ALTERNATING DIRECITON METHOD OF MULTIPLIERS

ADMM is a well-established technique for distributed convex optimization that decomposes a global problem into local subproblems while maintaining coordination through shared global variables and constraints [4]. These are normally problems of the form in (5), where each local function $f_i$ is optimized independently using an augmented Lagrangian, which is formulated by appending the global constraint (5b) into the objective function as shown in (6) which in the NetMEM problem corresponds to constraint (4). This relaxes the global constraints by assigning penalties to promote solutions that respect such constraints [4].

$$minimize \ f(x) = \sum_{i \in N} f_i(x_i) \quad (5a)$$

$$subject \ to \ \sum_{i \in N} A_i x_i = b \quad (5b)$$

$$L(x, y) = \sum_{i \in N} f_i(x) + \sum_{i \in N} y^T (A_i x_i - b) + \frac{\rho}{2} \| \sum_{i \in N} (A_i x_i - b) \|_2^2 \quad (6)$$

The ADMM algorithm iterations must be formulated in an alternating and sequential manner because the augmented Lagrangian (6) is not conditionally independent since the Lagrange multipliers are composed of all global variables. Thus, a simultaneous optimization of each system would give different, usually worse results than the optimization of all subsystems sequentially [4], [18]. This means that within the same iteration, the individual system must be optimized in a predefined order, with each subsystem utilizing the most up to date information from the previously optimized system. Following this approach, the ADMM iterations for a network example of three systems would be formulated as in (7).

$$x_1^{k+1} = argmin \ L(x_1, x_2^k, x_3^k, y^k)$$
$$x_2^{k+1} = argmin \ L(x_1^{k+1}, x_2, x_3^k, y^k) \quad (7)$$
$$x_3^{k+1} = argmin \ L(x_1^{k+1}, x_2^{k+1}, x_3, y^k)$$

In addition to the variable updates of (7), Lagrange multipliers as well as the primal and dual residuals must be updated for every iteration, as defined in (8).

$$y^{k+1} = y^k + \rho(A_1 x_1^{k+1} + A_2 x_2^{k+1} + A_2 x_2^{k+1}) \quad (8a)$$

$$r^{k+1} = \sum_{i \in N} (A_i x_i^{k+1} - b) \quad (8b)$$

$$s^{k+1} = \sum_{i \in N} (A_i x_i^{k+1} - A_i x_i^k) \quad (8c)$$

Ideally ADMM converges when the primal and dual residuals reach zero. Therefore, a feasibility metric that combines both of these two parameters can be used as stopping criteria to measure how well the relaxed constraint is being met at every iteration [12]. This metric is defined in (9).

$$\varepsilon^k = \sqrt{\|r^{L^k}\|_2^2 + \|s^{L^k}\|_2^2} \quad (9)$$

Although convergence ideally occurs when both residuals approach zero, this does not guarantee proximity to the true optimal solution because the metric in (9) only reflects constraint satisfaction [12]. Moreover, ADMM performance depends heavily on the choice of the penalty parameter $\rho$ that is used in (6) and (8a), and improper value selections can slow convergence, produce suboptimal results, or even cause divergence [12]-[13].

While ADMM ensures convergence for convex continuous problems, mixed-integer problem (MIP) formulations like the NetMEM can introduce complexities in the problem that may yield unstable or suboptimal results, mainly due to the imperfect optimality gap obtained when solving MIPs at every ADMM iteration. Nonetheless, several studies have shown that ADMM can still produce near-optimal solutions for certain nonconvex and MIP formulations [14], [19], though this is not a general guarantee, and it is dependent on many factors such as initial conditions and proper hyperparameter value selections. These challenges highlight the importance of improving initialization and developing strategies that enhance robustness to penalty selection.

Therefore, this paper considers three versions of the ADMM algorithm explained in the following subsections: the standard ADMM, a relaxed ADMM, and an objective-based ADMM (OB-ADMM). All of these versions follow the general flowchart shown in Fig. 1, where the red diamond represents the chosen convergence criteria.

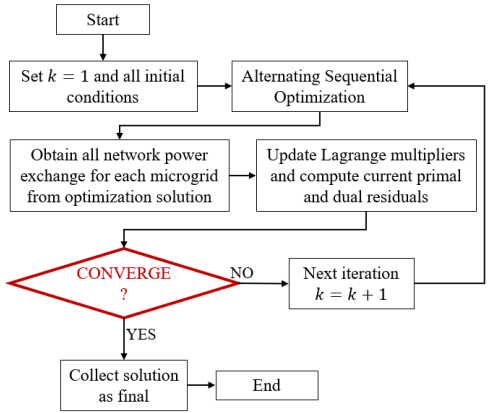

Fig. 1: Standard ADMM algorithm.

### A. Standard ADMM

The standard ADMM algorithm is the version that only checks the convergence of the feasibility metric to zero, which is a measure of the reduction in primal and dual residual, as determined by (9). This algorithm determines convergence once this metric drops below a predefined threshold, namely when $\varepsilon^k < \varepsilon_{th}$.

### B. Objective-Based ADMM

The OB-ADMM algorithm incorporates a couple more convergence criteria to check before making the decision to stop the ADMM iterations and collect the final solution. These are criteria that check for a decreased change in objective value as iterations progress, as well as the average solution feasibility to ensure $\varepsilon$ is less than this average as well as lower than $\varepsilon_{th}$ [15].

This objective-based approach modifies the standard ADMM to prevent early stopping that may lead to a feasible but suboptimal solution. The premise is that the objective value's rate of change begins to flatten as it approaches the optimal solution. Thus, once the average rate of change for the past $k_s$ iterations is lower than a predefined threshold $\beta$ then the algorithm begins to inspect feasibility as well to see if the current $\varepsilon$ is less than the average feasibility in the previous $k_s$ iterations as well as less than $\varepsilon_{th}$ [15]-[16]. The hyperparameters $\beta$ and $k_s$ can be freely selected by the user depending on the accuracy desired.

A summary of the convergence criteria for the standard ADMM and OB-ADMM are shown in Table I, which are the strategies that are used in the red diamond block of Fig. 1. This summary of criteria shows how the OB-ADMM is basically an enhancement to the standard ADMM by considering more criteria in addition to feasibility convergence.

TABLE I: Convergence criteria summary for standard and objective-based approaches for deciding a stopping point for ADMM.

| ADMM Strategy | Standard | Objective-Based |
|---|---|---|
| Hyperparameters | $\varepsilon_{th}$: Feasibility threshold | $\varepsilon_{th}$: Feasibility threshold<br>$\beta$: Avg. obj. value rate of change threshold.<br>$k_s$: Iteration offset |
| Convergence Criteria | Is $\varepsilon^k < \varepsilon_{th}$? | Is Avg. obj. val. rate of change $< \beta$?<br>And $\varepsilon^k <$ avg $\varepsilon$?<br>And $\varepsilon^k < \varepsilon_{th}$? |

### C. Relaxed ADMM

The relaxed ADMM algorithm implements a modified version of the local optimization problems for each MG in the network in which the binary decision variables of constraints such as those in (2) are relaxed as continuous variables with a range between 0 and 1, allowing them to take on fractional values. These relaxed versions of the original optimization problems are used in the ADMM decentralized operation until a certain feasibility threshold of $\varepsilon_{th0}$ is reached, which is larger than the actual threshold $\varepsilon_{th}$ that is used. Once this larger threshold is reached, the optimization problem is returned to its mixed-integer original version and the ADMM iterations are resumed using the last Lagrange multipliers, residuals, and power exchange values determined by the last iteration of the relaxed problem. This is known as the refinement step [14]. A flowchart depicting this relaxed version of ADMM is shown in Fig. 2.

### D. Adaptive Penalty Heuristic

ADMM can include a step in which dynamic updates to the penalty term are made depending on convergence performance of the primal and dual residuals. As a reference, the primal and dual residuals for this NetMEM problem are defined in (10), based on the global constraint (7) involving the network power exchanges.

$$r_{m,n,t}^{L^{k+1}} = P_{m,n,t}^{N+^{k+1}} - P_{n,m,t}^{N-^{k+1}} \quad (10a)$$

$$s_{m,n,t}^{L^{k+1}} = \left(P_{m,n,t}^{N+^{k+1}} - P_{n,m,t}^{N-^{k+1}}\right) - \left(P_{m,n,t}^{N+^{k}} - P_{n,m,t}^{N-^{k}}\right) \quad (10b)$$

Based on these metrics, the following heuristic in (11) can be implemented to dynamically update the parameter [6], [12].

$$\rho^{k+1} = \begin{cases} \tau\rho^k, & \text{if } \left\|r^{L^k}\right\|_2 > \mu \left\|s^{L^k}\right\|_2 \\ \frac{\rho^k}{\tau}, & \text{if } \left\|s^{L^k}\right\|_2 > \mu \left\|r^{L^k}\right\|_2 \\ \rho^k, & \text{otherwise} \end{cases} \quad (11)$$

This scheme has been found effective for a variety of problems; however, it is important to note that in some instances it may affect convergence instead of improving it due to scaling properties impacting the residuals. Moreover, the most common values for $\mu$ and $\tau$ are 10 and 2, respectively [12]. This adaptive penalty step is incorporated into all three ADMM versions considered in this paper to test and compare their convergence performance and final solution optimality results against their fixed-penalty counterparts.

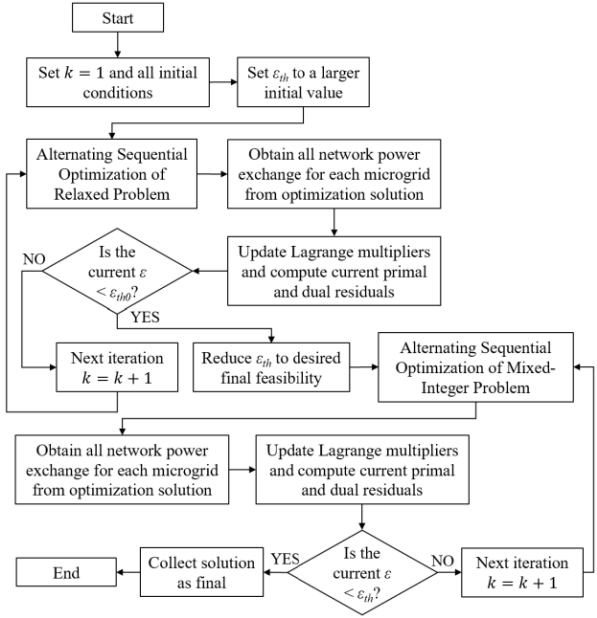

Fig. 2: Relaxed ADMM algorithm.

IV. CASE STUDIES

Three test systems comprising networks of three, four, and six MGs are used to evaluate the three ADMM formulations described earlier. Each MG has distinct net load and main grid price profiles to emulate heterogeneous conditions across the network, in addition to scalability assessment by using MG networks varying sizes. All tests utilize a feasibility threshold $\varepsilon_{th} = 0.01$ for convergence checks for all ADMM algorithm versions, and hyperparameters $\beta = 0.001$ and $k_s = 25$ are used for OB-ADMM. The penalty parameter is varied as $\rho \in$ [0.00001, 0.0001, 0.001, 0.01, 0.1, 1 10, 100], with adaptive penalty versions using these as initial values $\rho_0$ in the update scheme of (11).

Table II summarizes the optimality results for all three of the test cases using the three different ADMM variants and penalty selections, where optimality is measured by the percentage deviation from the centralized benchmark results. These numerical results clearly demonstrate that the OB-ADMM consistently achieves the smallest optimality gap across nearly all cases, confirming its robustness to penalty selection and ability to avoid premature stopping. For smaller networks, such as three and four MGs, OB-ADMM combined with the adaptive penalty heuristic yields solutions that are practically identical to the centralized benchmark for a wide range of penalty values, highlighting the improved stability and convergence reliability of this method.

As the network scale increases, such as for the six-MG test case, the adaptive penalty heuristic occasionally causes divergence for certain values of $\rho$ due to scaling effects in residual balancing, which is consistent with the claims in [12]. Moreover, when convergence occurs, the improvements over fixed-penalty configurations are marginal. In contrast, the relaxed ADMM offers mild gains in solution optimality compared to the standard formulation, but these improvements are limited relative to those of the OB-ADMM. Fig. 3 illustrates the optimality results for the four-MG test case under $\rho = 0.001$ and 0.01, confirming that across all tested penalties: Standard ADMM < Relaxed ADMM < OB-ADMM.

TABLE II: Optimality results in terms of the percentage difference from centralized benchmark solutions under different ADMM versions.

| Test Case | Penalty | Standard ADMM | | Relaxed ADMM | | OB-ADMM | |
|---|---|---|---|---|---|---|---|
| | | Fixed $\rho$ | Variable $\rho$ | Fixed $\rho$ | Variable $\rho$ | Fixed $\rho$ | Variable $\rho$ |
| Case 1 (3 microgrids) | 0.0001 | 0.0387 | 0.0387 | 0.0207 | 0.0000 | 0.0000 | 0.0000 |
| | 0.001 | 1.3728 | 1.3728 | 0.0742 | 0.0326 | 0.0000 | 0.0000 |
| | 0.01 | 1.6703 | 1.6703 | 1.6040 | 1.6040 | 0.0083 | 0.0000 |
| | 0.1 | 1.7001 | 1.7001 | 1.6957 | 1.6946 | 0.0462 | 0.0000 |
| | 1 | 1.7031 | 1.7031 | 1.7031 | 1.7031 | 0.3008 | 0.0000 |
| | 10 | 1.7034 | 1.7034 | 1.7034 | 1.7034 | 1.7031 | 0.0000 |
| Case 2 (4 microgrids) | 0.0001 | 0.0515 | 0.0000 | 0.0000 | 0.0000 | 0.0000 | 0.0000 |
| | 0.001 | 2.2896 | 0.1236 | 1.2526 | 0.0561 | 0.0000 | 0.0000 |
| | 0.01 | 3.4579 | 3.4369 | 3.3313 | 0.0498 | 0.0773 | 0.0000 |
| | 0.1 | 3.5317 | 3.5275 | 3.5275 | 3.5275 | 0.1460 | 0.0000 |
| | 1 | 3.5416 | 3.5416 | 3.5407 | 3.5407 | 0.3171 | 0.0000 |
| | 10 | 3.5422 | 3.5422 | 3.5422 | 3.5422 | 3.5417 | 0.0000 |
| Case 3 (6 microgrids) | 0.0001 | 0.0000 | DIVERGED | 0.0000 | DIVERGED | 0.0000 | DIVERGED |
| | 0.001 | 0.0000 | DIVERGED | 0.0000 | 0.0000 | 0.0000 | DIVERGED |
| | 0.01 | 0.0300 | 0.0251 | 0.0295 | 0.0000 | 0.0281 | 0.0251 |
| | 0.1 | 0.1078 | 0.1121 | 0.1146 | 0.0882 | 0.0334 | DIVERGED |
| | 1 | 0.1263 | 0.1245 | 0.1268 | 0.1268 | 0.1238 | DIVERGED |
| | 10 | 0.1267 | 0.1267 | 0.1267 | 0.1267 | 0.1265 | 0.1255 |

While the OB-ADMM provides undeniable improvements in optimality across a wide range of penalties and network sizes, it requires more iterations before meeting its convergence conditions, hence taking longer computation time. This is reflected in Table III which summarizes the total number of iterations taken for Test Case 2 for each ADMM approach under different values of $\rho$. Nonetheless, this extra computing time corresponds directly to improved optimality due to the prevention of early stopping, which is not achieved with the standard or the relaxed ADMM. Also, the hyperparameters $\beta$ and $k_s$ can be fine-tuned to balance accuracy and computational effort, offering a clear trade-off decision the user can make, in contrast to the uncertain nature of the penalty hyperparameter $\rho$ whose effect on convergence is less predictable.

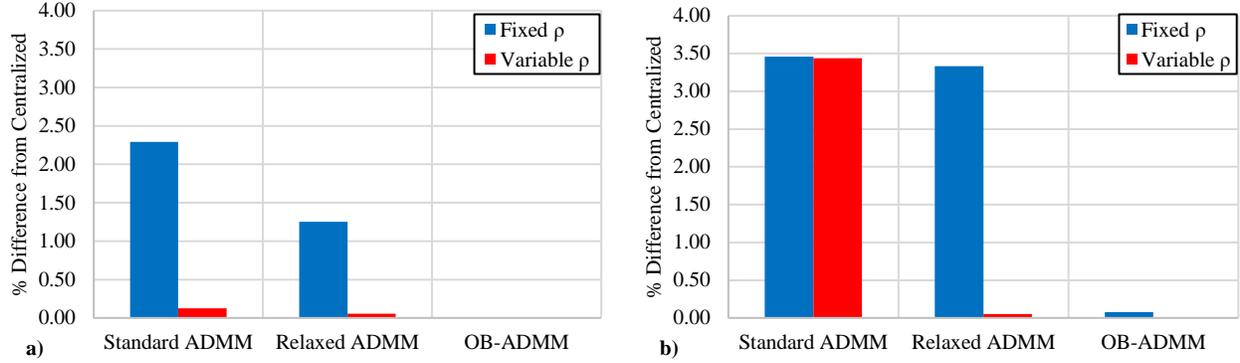

Fig. 3: Optimality results for Test Case 2 of four microgrids using different ADMM methods and penalties of a) $\rho = 0.001$ and b) $\rho = 0.01$.

TABLE III: Number of iterations by each ADMM method for Test Case 2.

| Penalty ($\rho$) | | 0.001 | 0.01 |
|---|---|---|---|
| Standard ADMM | Fixed ($\rho$) | 6 | 4 |
| | Variable ($\rho$) | 14 | 4 |
| Relaxed ADMM | Fixed ($\rho$) | 11 | 10 |
| | Variable ($\rho$) | 24 | 29 |
| OB-ADMM | Fixed ($\rho$) | 385 | 69 |
| | Variable ($\rho$) | 1463 | 80 |

## V. CONCLUSIONS

This paper evaluated three different ADMM formulations, Standard ADMM, Relaxed ADMM, and Objective-Based ADMM, for decentralized networked microgrid energy management, emphasizing robustness to the penalty parameter $\rho$ and the effect of an adaptive penalty heuristic based on residual balancing. Case studies on networks of three, four, and six microgrids showed that OB-ADMM consistently achieved higher final solution optimality than the other two approaches by incorporating convergence criteria that evaluate the objective value's rate of change in addition to solution feasibility, effectively avoiding premature algorithm termination at suboptimal solutions.

The adaptive penalty heuristic improved convergence for smaller networks but introduced scalability challenges in larger ones, occasionally leading to divergence. Although OB-ADMM requires more iterations and thus longer computation times, its tunable hyperparameters allow users to balance accuracy and computational efficiency more intuitively than tuning the penalty parameter $\rho$, whose influence on ADMM convergence remains unclear and problem-dependent. Overall, OB-ADMM demonstrated the best trade-off between robustness, accuracy, and user control among the evaluated formulations.